\documentclass{elsart}

\usepackage{amssymb}
\usepackage{comment}
\usepackage{graphicx}
\usepackage{natbib}

\newcommand{\fd}{\mbox{d}} 
\newcommand{\pd}{\partial}   

\begin{document}


\begin{frontmatter}

\title{The Role of Angular Momentum Transport in Establishing the Accretion Rate--Protostellar Mass Correlation}

\author{Alexander L. DeSouza} \&
\ead{alexander.desouza@gmail.com}
\author{Shantanu Basu}
\ead{basu@uwo.ca}

\address{Department of Physics \& Astronomy, The University of Western Ontario \\ 1151 Richmond Street, London, ON, Canada, N6A 3K7}

\begin{abstract}
We model the mass accretion rate $\dot{M}$ to stellar mass $M_*$ correlation that has been inferred from observations of intermediate to upper mass T Tauri stars---that is $\dot{M} \propto M_*^{1.3 \pm 0.3}$. We explain this correlation within the framework of quiescent disk evolution, in which accretion is driven largely by gravitational torques acting in the bulk of the mass and volume of the disk. Stresses within the disk arise from the action of gravitationally driven torques parameterized in our 1D model in terms of Toomre's $Q$ criterion. We do not model the hot inner sub-AU scale region of the disk that is likely stable according to this criterion, and appeal to other mechanisms to remove or redistribute angular momentum and allow accretion onto the star. Our model has the advantage of agreeing with large-scale angle-averaged values from more complex nonaxisymmetric calculations. The model disk transitions from an early phase (dominated by initial conditions inherited from the burst mode of accretion) into a later self-similar mode characterized by a steeper temporal decline in $\dot{M}$. The models effectively reproduce the spread in mass accretion rates that have been observed for protostellar objects of $0.2\,\mbox{M}_\odot \le M_* \le 3.0\,\mbox{M}_\odot$, such as those found in the $\rho$ Ophiuchus and Taurus star forming regions. We then compare realistically sampled populations of young stellar objects produced by our model to their observational counterparts. We find these populations to be statistically coincident, which we argue is evidence for the role of gravitational torques in the late time evolution of quiescent protostellar disks.
\end{abstract}

\begin{keyword}
accretion; accretion disks; hydrodynamics; stars: formation; stars: protostellar disks
\end{keyword}

\end{frontmatter}


\section{Introduction}
\label{sec:introduction}

Protostellar disks are a ubiquitous outcome of the rotating collapse of dense molecular cloud cores in the standard paradigm of low-mass star formation \citep[e.g.,][]{terebey1984,shu1987}. Their existence has been confirmed around young stellar objects across a broad range in mass---from objects in the brown dwarf regime, to those with masses of up to $2\mbox{--}3\,\mbox{M}_\odot$ \citep[e.g.,][]{beckwith1990,andrews2005}---as well as in a wide variety of star forming environments \citep[e.g.,][]{lada1984,o'dell1994,mccaughrean1996}.

Numerical simulations of collapsing cloud cores reveal that disks can form within ${\sim}10^4\,\mbox{yr}$ from the onset of core collapse \citep{yorke1993,hueso2005}. These early so-called Class 0 systems are difficult to study observationally as they are still embedded within their progenitor cloud cores \citep{andre1993}. Numerical simulations \citep[e.g.,][]{vorobyov2005b,vorobyov2006,vorobyov2010,vorobyov2015} suggest that the earliest periods (${\sim}0.5\,\mbox{Myr}$) of disk formation are rather tumultuous, as infall from the parent cloud core induces gravitational instability--driven mass accretion. Depletion of the gas reservoir by this mechanism then gives way to a much more quiescent period of accretion in which gravitational torques act to transport mass inward while transporting angular momentum outward \citep{gammie2001,lodato2004,vorobyov2007}. Indeed, the subsequent Class I and II phases are respectively marked by a decline in the rate of accretion from the surrounding natal environment, and its eventual cessation \citep{vorobyov2005a}. Hence, it is during the Class II phase, once the central star is optically visible, that the disk properties are most easily amenable to observational investigation.

One result to emerge from observational studies of young stellar objects and their disks is the correlation between protostellar mass $M_*$ and the inferred accretion rate $\dot{M}$ from the disk, for which the power law exponent is typically estimated to be $\beta \sim 1.5\mbox{--}2.0$ \citep[e.g.,][]{muzerolle2005,herczeg2008,rigliaco2011}. Although this correlation appears to hold across multiple orders of magnitude in both $M_*$ and $\dot{M}$, fitting the accretion rates of brown dwarfs and T Tauri stars together may be misleading. In the brown dwarf regime, as well as for low mass T Tauri stars (i.e., those objects with mass $M_* < 0.2\,\mbox{M}_\odot$), a least squares fit yields $\beta = 2.3 \pm 0.6$. For intermediate and upper mass T Tauri stars ($M_* > 0.2\,\mbox{M}_\odot$), the equivalent fit yields a value for $\beta$ of $1.3 \pm 0.3$; suggestive that different physical mechanisms may be responsible for accretion across the sequence of protostellar masses \citep{vorobyov2008}.

Studies by \citet{alexander2006} and \citet{hartmann2006} have sought to explain the $\dot{M}\mbox{--}M_*$ scaling in the context of viscous models for the disk evolution, wherein the turbulent viscosity has ad hoc spatial dependence of the form $\nu \propto r^\xi$. \citet{dullemond2006} link the disk evolution to the properties of the parent cloud core, providing a self-consistent basis for the results of their study. However, their models require that the ratio of rotational to gravitational energy be uniform across all cloud core masses. \citet{rice2009} have even attempted to (weakly) incorporate the additional effects of magnetic fields (in high temperature regions of the disk) in quasi--steady state models, but were also unable to fully account for the observed correlation.

In this paper we present a study of the quasi--steady state evolution of viscous circumstellar disks surrounding young stellar objects, following the cessation of mass accretion onto the protostar-disk system (definitively Class II objects). These disks inherit initial conditions roughly consistent with the results of numerical simulations of the earlier burst phase \citep[e.g.,][]{vorobyov2005b,vorobyov2006,vorobyov2010,vorobyov2015}, and undergo diffusive evolution wherein angular momentum redistribution is driven by self-gravity, which we parameterize in terms of an effective kinematic viscosity \citep[following][]{lin1987}. We add to this a simplified argument for angular momentum conservation that correlates disk size with protostellar mass at the start of our simulations. With these assumptions, we are able to reproduce many features of the observed correlation between $\dot{M}$ and $M_*$ for young protostellar systems.

Recent observations of disks using near-infrared polarization imaging \citep{liu2016} have found that disks around four recently outbursting (FU Ori) sources have large-scale (hundreds of AU) spiral arms and arcs that are consistent with models of gravitational instability. Added to previous near-infrared detections of spiral structure in smaller disks \citep[e.g.,][]{hashimoto2011, muto2012,grady2013}, there is a growing realization that meaningful spiral structure, arcs, and gaps exist in Myr-old disks (see the review by \citet{tamura2016} on the SEEDS survey by the Subaru telescope). New efforts are being made to use gravitational instability driven disk evolution models to predict the near-infrared scattered light patterns as may be seen by the Subaru or Gemini telescopes, or the millimeter dust emission patterns that may be seen with the ALMA telescope \citep{dong2016}. Furthermore, numerical simulations are also being extended to include long-term residual infall from the molecular cloud to the disk (even after the parent cloud core may have dissipated), which may be needed to keep gravitational instability active after several Myr \citep{vorobyov2015b,lesur2015}. Our model in this paper studies gravitational torque driven evolution in a simplified manner. It does not however include residual mass infall from the cloud, which may be a subject of future work.

In this paper we seek to characterize the bulk of transport within the disk through the action of gravitational torques, in the same spirit as the models of e.g., \citet{armitage2001} and \citet{zhu2009,zhu2010}. Our aim is to explain the global behavior of disks in which the mass accretion rate is predominantly set by the action of gravitational torques acting through most of the disk. Other accretion mechanisms may be necessary in the innermost sub-AU regions of the disk, possibly introducing short-term time variability. The above studies typicaly invoke the magnetorotational instability \citep{balbus1991} as the transport mechanism in the hot inner disk, however it is worthwhile to keep in mind that the region $0.1\mbox{--}1.0\,\mbox{AU}$ from the star is generally thought to be the outflow driving zone \citep[e.g.,][]{garcia2001,krasnopolsky2003} from which significant amounts of angular momentum and mass are carried away from the disk.


\section{Disk Model}
\label{sec:diskmodel}

We construct a model for the temporal evolution of self-gravitating, axisymmetric thin disks on a radial grid with logarithmic spacing, and consisting of $N = 256$ annular elements. Discretization of the radial grid allows us to write the relevant partial differential equations as sets of ordinary differential equations, with one equation for each coordinate position in $r$. The spatial derivatives are approximated using second-order accurate central differencing. Integration of the system through time is handled using a variable order Adams-Bashforth-Moulton solver \citep[e.g.,][]{shampine1994book}.

\subsection{Viscous Evolution of an Axisymmetric Thin Disk}
\label{subsec:thindiskeaccretion}

Combining together the fluid equations for mass and momentum conservation yields a diffusion-like equation that governs the temporal evolution of the disk surface mass density $\Sigma(r,t)$ \citep[e.g.,][]{lyndenbell1974,pringle1981}:
\begin{equation}
\label{eqn:diskevolutioneqn}
\frac{\pd \Sigma}{\pd t}
=
-
\frac{1}{r} \frac{\pd}{\pd r}
\left[
\left(
\frac{\pd}{\pd r} r^2 \Omega
\right)^{-1}
\frac{\pd}{\pd r}
\left(
\nu r^3 \Sigma \frac{\pd \Omega}{\pd r}
\right)
\right],
\end{equation}
where $\Omega(r,t)$ is the disk angular frequency (obtained assuming centrifugal balance), and $\nu$ is the effective kinematic viscosity (detailed in Section \ref{subsec:viscosity}).

A precise determination of $\Omega$ requires a thorough accounting of the contribution to the gravitational potential made by the disk itself, which can be calculated explicitly using the elliptic integral of the first kind \citep[e.g.,][]{binney2008book}. However, the central point-mass dominates the system's gravitational potential, with the contribution from the disk increasing $\Omega$ only slightly. For the sake of computational convenience we thus adopt a simplified procedure by approximating the total gravitating mass at a radius $r$ to be
\begin{equation}
\label{eqn:selfgravitatingdiskmass}
M(r,t) = M_*(t) + 2 \pi \int_{\rm r_{in}}^r \Sigma \, r' \, \fd r',
\end{equation}
in which $r_{\rm in}$ denotes the innermost radius of the simulation domain (and the assumed disk inner edge).

The action of (\ref{eqn:diskevolutioneqn}) is to transport material within the disk to ever smaller radii, while a small fraction of disk material is simultaneously transported to larger radii, thereby preserving the system's total angular momentum. For these simulations, the disk edge $r_{\rm edge}$ is always $\ll r_{\rm out}$, the computational domain's outer boundary. Thus, material that exits the simulation can only do so through $r_{\rm in}$. We impose a free \textbf{outflow} boundary condition there, and any material crossing $r_{\rm in}$ is assumed to be accreted onto the central protostar, which we model as a point mass.

\subsection{Viscosity}
\label{subsec:viscosity}

Temporal evolution of the disk is governed by the viscous stresses acting on the disk material. These stresses are typically subsumed into a dimensionless parameter $\alpha$ that characterizes the efficiency of angular momentum transport. \citet{shakura1973} developed the most commonly invoked prescription of this kind, proposing a turbulent kinematic viscosity of the form
\begin{equation}
\nu = \alpha v \ell,
\end{equation}
which is the product of the turbulent velocity $v$ and the size $\ell$ of the largest eddies in the turbulent pattern. As turbulence is quickly dissipated by shocks in a highly supersonic flow, the turbulent velocity is often taken to be roughly equal to the local sound speed of the disk medium, $c_{\rm s}$. An upper limit to the size of the largest eddies that form can similarly be argued to be roughly equal to the disk half-thickness $H$; hence
\begin{equation}
\label{eqn:alphaviscosity}
\nu = \alpha \frac{c_{\rm s}^2}{\Omega},
\end{equation}
where we have used $H = c_{\rm s} / \Omega$, and imposed that the disk be everywhere in vertical hydrostatic equilibrium. Taken together, these arguments imply that $\alpha$ should be less than unity. Determinations based on measurements of the accretion rates from disks surrounding young stellar objects in the Taurus complex suggest that $\alpha \sim 10^{-2}$ \citep{hartmann1998}. However, if and why $\alpha$ should be spatially uniform and constant in time remains unclear.

One mechanism that may act as an effective $\alpha$-viscosity is the magnetorotational instability (MRI). \cite{balbus1991} have demonstrated that a weak magnetic field can make an otherwise hydrodynamically stable disk become unstable to the MRI. Simulations in ideal magnetohydrodynamics (MHD) find that the nonlinear outcome of MRI can amplify and sustain turbulence that can then lead to angular momentum transport \citep{brandenburg1995,hawley1996}. However, in a largely neutral medium (such as in a protostellar disk) the ionization fraction must be large enough for the neutral-ion collision frequency to be greater than the local epicyclic frequency \citep{blaes1994}. Many studies have been performed to determine in which regions of protostellar disks the MRI may be active \citep{gammie1996,igea1999,sano2000,fromang2002,cleeves2013}. For example, \citet{igea1999} find that the MRI can be active at distances greater than 5 AU from the central star, while \citet{cleeves2013} finds that the disk midplane is largely inactive to the MRI after accounting for the effect of stellar winds and magnetic mirroring of cosmic rays, in addition to using a different critical electron fraction for the magnetic coupling. Simulations of the MRI with non-ideal MHD are very sensitive to ionization structure \citep[e.g.,][]{fleming2003}. Hence the determination of an effective $\alpha$ for the MRI in local simulations, let alone for a global disk model or for different evolutionary stages, is hardly established. \citet[][and references therein]{pp4stone2000coll} suggest a value for $\alpha$ anywhere in the range of $10^{-3}\mbox{--}0.5$. When using $\alpha$ as a comparator in this study we use $\alpha = 0.01$, based on observational constraints (Hartmann et al., 1998).

Gravitational torques may represent an alternative source for angular momentum transport in cold and/or massive disks. Using self-consistent cooling, \citet{boley2006} showed that an $\alpha$ prescription based on the gravitational instability agrees with the \citet{gammie2001} description very well, which assumes that the viscous heating is locally balanced by the cooling. \citet{cossins2009} have also used smoothed particle hydrodynamics to show that these disks often possess tightly wound spiral arms that can be approximated with a local treatment. Here we consider a perturbation in an otherwise axisymmetric thin disk, which has the form of an annulus of width ${\Delta}r$ and increased local mass ${\Delta}m$ (e.g., the formation of a spiral arm). The growth condition for a perturbation depends on whether its self-gravity is greater than the tidal acceleration acting on it. That is,
\begin{equation}
\frac{G {\Delta}m}{{\Delta}r^2} \sim \pi G \Sigma > \frac{G M_*}{r^2}\frac{{\Delta}r}{r}.
\end{equation}
A natural length scale thus emerges, in excess of which perturbations of this nature are stabilized by their rotation,
\begin{equation}
{\Delta}r \sim \pi G \Sigma \left( \frac{G M_*}{r^3} \right)^{-1} = \frac{\pi G \Sigma}{\Omega^2}.
\end{equation}
Furthermore, with the assumption of vertical hydrostatic equilibrium, the disk self-gravity is supported by gas pressure in the vertical direction. This additional constraint implies ${\Delta}r$ must be at least larger than the disk half-thickness $H$,
\begin{equation}
{\Delta}r \geq H = \frac{c_{\rm s}}{\Omega}.
\end{equation}
\citet{toomre1964} originally formulated these arguments, summarizing the instability criterion as:
\begin{equation}
\label{eqn:toomresqcriterion}
Q \equiv \frac{c_{\rm s} \Omega}{\pi G \Sigma} < 1.
\end{equation}
This condition can additionally be rephrased in terms of the disk mass. Multiplying equation (\ref{eqn:toomresqcriterion}) by the square of the disk outer radius, $r_{\rm out}^2$, and then approximating the disk mass as $M_{\rm disk} \sim \pi r_{\rm out}^2 \Sigma$, Toomre's $Q$ criterion implies
\begin{equation}
\frac{M_{\rm disk}}{M_*} > \frac{H}{r}.
\end{equation}
From this statement one can conclude that provided the disk is thin, even a relatively low-mass disk will exhibit the effects of self-gravity. Disks of this variety may not be uncommon, possibly forming during the earliest stages of star formation \citep[e.g.,][]{eisner2006}. The importance of self-gravity in providing an effective way of redistributing angular momentum at the earliest stages of star formation has been recently recognized \citep{vorobyov2005b,hartmann2006,vorobyov2006}.

\citet{lin1987} noted that gravitational torques could be parameterized as an effective kinematic viscosity, constructed dimensionally using the length scales arising from Toomre's analysis: the maximum size of the region over which angular momentum is transferred being roughly $G \Sigma / \Omega^2$, together with a time-scale of approximately $\Omega^{-1}$, produces an effective kinematic viscosity of the form
\begin{equation}
\label{eqn:lp1987viscosity}
\nu \sim \left( \frac{G \Sigma}{\Omega^2} \right)^2 \frac{1}{\Omega^{-1}} \sim Q^{-2} \, \frac{c_{\rm s}^2}{\Omega}.
\end{equation}
This is clearly analogous to the standard $\alpha$ prescription of \citet{shakura1973} with $\alpha = Q^{-2}$ (see equation (\ref{eqn:alphaviscosity})). Additionally, by using equation (\ref{eqn:toomresqcriterion}) we can also write
\begin{equation}
\label{eqn:convenientnu}
\nu = r^6 \Sigma^2 \Omega / M^2.
\end{equation}
Hence, a convenient feature of parameterizing the gravitational torques via equation (\ref{eqn:convenientnu}) is that we need not explicitly evaluate the energy equation during the disk evolution. This allows us to study the evolution of the disk while circumventing the complex issues related to the disk thermodynamics. Nevertheless, to validate the consistency of our results with existing 1D simulation work \citep[e.g.,][]{armitage2001,rice2009} as well as higher dimensional models, such as the 2+1D models of \citet{vorobyov2006,vorobyov2010,vorobyov2015}, we do follow the temperature evolution of the disk implicitly.

In calculating the temperature of the disk we assume it evolves isothermally up to some critical density $\Sigma_{\rm crit}$, and in a polytropic manner thereafter. The critical surface mass density at which this transition occurs is $\Sigma_{\rm crit} = 36.2\,\mbox{g\,cm}^{-2}$, and corresponds to a critical volume density of $n_{\rm crit} \sim 10^{11}\,\mbox{cm}^{-3}$ for a gas disk in vertical hydrostatic equilibrium at $T = 10\,\mbox{K}$ \citep{larson2003}. Matching the isothermal and non-isothermal regimes, the effective vertically integrated gas pressure as a function of surface mass density can therefore be expressed as \citep{vorobyov2006}
\begin{equation}
P = c_{\rm s,0}^2 \Sigma + c_{\rm s,0}^2 \Sigma_{\rm crit} \left( \frac{\Sigma}{\Sigma_{\rm crit}} \right)^{\gamma}.
\end{equation}
Here, $c_{\rm s,0}$ is the sound speed corresponding to a medium of predominantly molecular hydrogen (with an admixture of helium) that is isothermal with $T = 10\,\mbox{K}$. For the ratio of the specific heats we used $\gamma = 5/3$.

The gas temperature, via the ideal gas equation of state $P = \Sigma k_{\rm B} T / \mu m_{\rm p}$, is then simply
\begin{equation}
T = \frac{ c_{\rm s,0}^2 \mu m_{\rm p} }{ k_{\rm B} } \left[ 1 + \left( \frac{\Sigma}{\Sigma_{\rm crit}} \right)^{\gamma-1} \right],
\end{equation}
where $\mu$ is the mean molecular mass (which we take to be 2.36), $m_{\rm p}$ is the proton mass, and $k_{\rm B}$ is Boltzmann's constant.

Angular momentum transport by gravitational instability (or more appropriately, gravitational torques, as being described herein) has been shown to remain effective in the regime of $Q \gtrsim 1$ \citep{vorobyov2006,vorobyov2007,vorobyov2009a,vorobyov2009b}. In their 2006 study specifically, it was found that radial profiles of the Toomre $Q$ parameter were both near-uniform and noticeably larger during periods of quiescence in the mass accretion rate, and never falling below $2.5\mbox{--}3.0$ during this phase. Although the persistence of gravitational instabilities is not usually expected for these values of $Q$, their 2+1D simulations revealed that weak spiral structures formed early on in the formation of the disk were then sustained by a swing amplification mechanism\citep{vorobyov2007}. In fact, strong observational evidence for spiral structure has actually been found in several-million year old disks such as HD 100546 \citep{grady2001}, AB Aurigae \citep{fukagawa2004}, and HD 135344B \citep{muto2012}, all in support of this conjecture.


\section{Results}
\label{sec:results}

We investigate the temporal evolution of more than 200 initial protostar-disk configurations. The parameter space of our models include initial protostellar masses of $0.2\,\mbox{M}_\odot \leq M_* \leq 3.0\,\mbox{M}_\odot$, corresponding to the range of intermediate- to upper-mass T Tauri stars. This range is divided into intervals of $0.1\,\mbox{M}_\odot$ up to $1.1\,\mbox{M}_\odot$, and intervals of $0.2\,\mbox{M}_\odot$ thereafter. Each protostar harbors a disk that extends from $r_{\rm in} = 10^{-1}\,\mbox{AU}$ to an initial size $r_{\rm edge}$ that is $\ll r_{\rm out} = 10^4\,\mbox{AU}$, ensuring that the simulation's outer boundary has no influence on the disk evolution. We note that since we model the late-time quiescent evolution of the disk, the time $t = 0$ in our simulations represents the state of a disk that is already $\gtrsim 10^5\,\mbox{yr}$ old \citep[e.g., see Figure 1 in][]{vorobyov2007}. Our simulations end after an additional $2{\times}10^6\,\mbox{yr}$, congruent with observational estimates of disk lifetimes in the post-embedded phase \citep[see][and references therein]{williams2011}.

\subsection{Initial Conditions}
\label{subsec:initialconditions}

\begin{figure}
\centering
\includegraphics[width=\columnwidth]{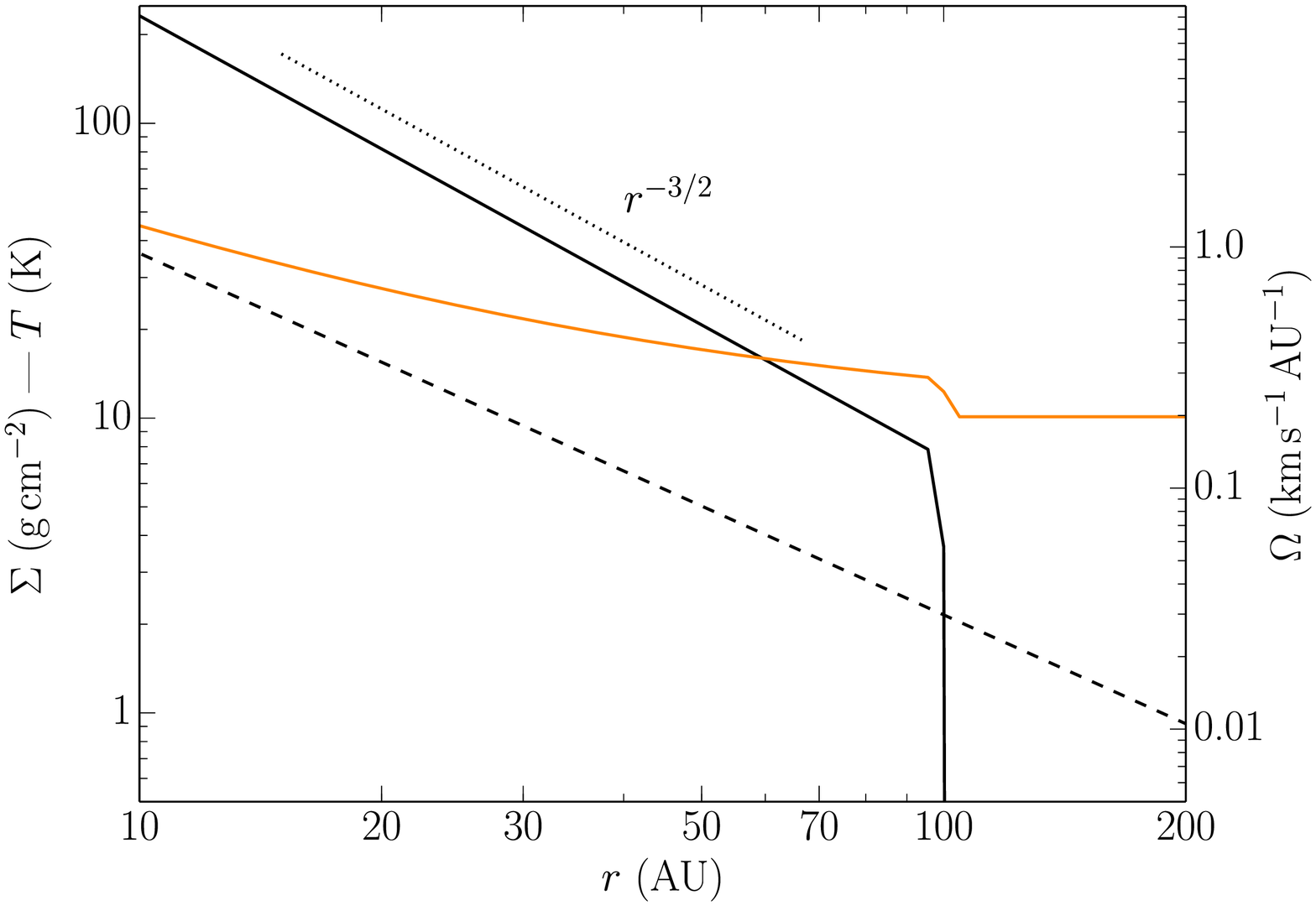}
\includegraphics[width=\columnwidth]{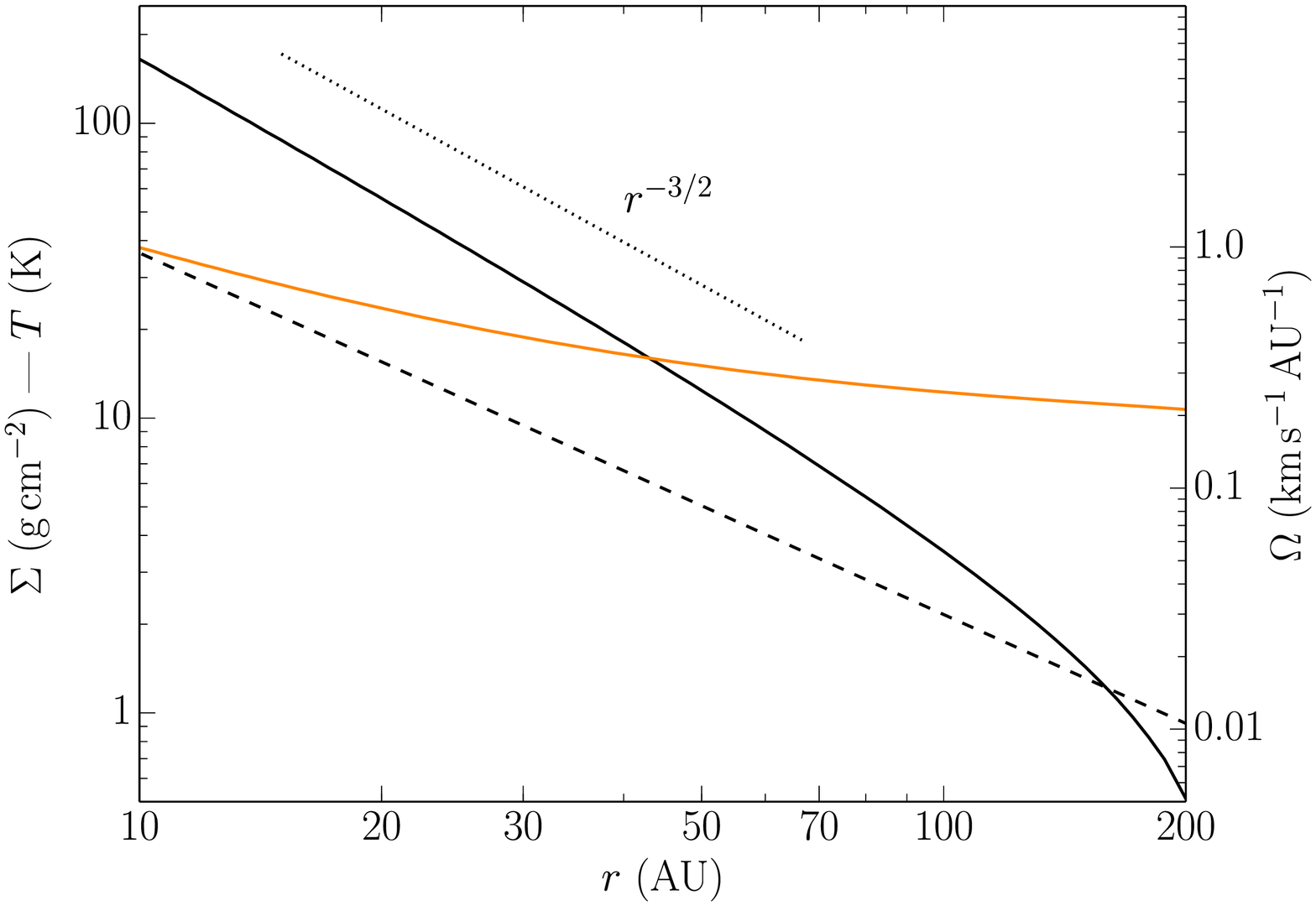}
\caption{
Radial profiles of disk surface mass density $\Sigma$ (solid black lines), angular velocity $\Omega$ (dashed black lines), and temperature $T$ (solid orange lines), at the beginning of the simulation ($t = 0$; top), and at its end ($t = 2 \times 10^6\,\mbox{yr}$; bottom). The protostar's initial mass is $1.0\,\mbox{M}_\odot$; its associated disk, $0.1\,\mbox{M}_\odot$. The correlation between surface mass denisty with radius as $\Sigma \propto r^{-3/2}$ is also shown to guide the eye (dotted black line).
}
\label{fig:diskevolutionexample}
\end{figure}

We presume that each protostar-disk system is formed as a result of the collapse of a rotating prestellar core. Numerical simulations of this process \citep[e.g.,][]{vorobyov2006,vorobyov2010,vorobyov2011,vorobyov2015} show that a disk formed in this manner undergoes an initial burst mode of accretion driven by gravitational instabilities. At the end of this phase, the disk settles into a more quiescent phase, which we model in this paper, and is characterized by a disk roughly $100\,\mbox{AU}$ in radius. We follow the numerical results of \citet{vorobyov2007,vorobyov2008} and adopt a prescription weakly correlating the disk mass (at $t = 0$) with the host protostar's mass, $M_{\rm disk} \propto M_*^{0.3}$. This produces disk mass fractions in the range of $5\%$ at the low mass end, to ${\sim}14\%$ for the highest mass stars in our study. Mass fractions of this order are greater than what is typically reported for disks with similar ages to those in our study \citep[e.g.,][]{andrews2005}. However, at face value, up to 10\% of the disk mass values reported by \citet{andrews2005} would still fall in this regime (see their Figure 10). Moreover, there is significant uncertainty in the disk masses estimated from dust emission. Dust grain growth to sizes $\gtrsim 1\,\mbox{mm}$, for example, decrease the millimeter wavelength opacity and allow for higher disk mass estimates \citep[e.g.,][]{d'alessio2001,williams2012}. \citet{dunham2014} have computed synthetic observations from which they argue that a combination of partial optical thickness and emission further from the Rayleigh-Jeans limit can lead to underestimates of the actual disk mass when assuming optically thin Rayleigh-Jeans emission from submillimeter observations (see their Figure 7).

Between $r_{\rm in}$ and $r_{\rm edge}$ the local surface mass density within the disk scales as $\Sigma \propto r^{-3/2}$, and is exponentially tapered thereafter. This is consistent with theoretical estimates for the minimum-mass solar nebula \citep{hayashi1981}, and the extrapolated radial profiles of observationally resolved disks \citep[e.g.,][]{hartmann1998}. For completeness we have also investigated the evolution of disks with two alternative initial profiles: $\Sigma \propto r^{-1}$ and $\propto r^{-2}$. However, the action of (\ref{eqn:diskevolutioneqn}) and (\ref{eqn:lp1987viscosity}) readjusts the surface mass density distribution of the disk to be $\propto r^{-3/2}$ within ${\sim}10^5\,\mbox{yr}$, rendering the initial distinction inconsequential for the subsequent evolution. Evolution toward $Q \approx 1$ in an environment with near Keplerian rotation and weak temperature variation will invariably lead to a profile $\Sigma \propto r^{-3/2}$, as can be seen from equation (\ref{eqn:toomresqcriterion}).

We estimate the scaling of the initial disk radial extent $r_{\rm edge}$ with central object mass $M_*$ as follows. A parcel of material initially located at a cylindrical radius $r$ that possesses a specific angular momentum $j$ and that falls in from the outermost mass shell of the cloud core, can be expected to land in the plane of the disk at $r_{\rm edge} = j^2 / G M_*$. Here, we have used a rough equality of core mass $M$ and final central object mass $M_*$. For a rotating and collapsing cloud core we expect surface mass density and rotation profiles $\Sigma \propto r^{-1}$ and $\Omega \propto r^{-1}$, respectively \citep{basu1997}. In this case, $j = \Omega r^2 \propto r$ and $M \propto r$, so that $j \propto M$. We therefore expect disk sizes to directly correlate with protostellar mass approximately as
\begin{equation}
\label{eqn:diskradiusrelation}
r_{\rm edge} \propto M_*,
\end{equation}
where we use the empirically motivated scale that $M_* = 1\,\mbox{M}_\odot$ corresponds to $r_{\rm edge} = 100\,\mbox{AU}$. We use this relation as a proxy for determining disk sizes in our models at $t = 0$.

In Figure \ref{fig:diskevolutionexample} we present snapshots of the radial profiles of the disk surface mass density $\Sigma$ (solid black lines), angular velocity $\Omega$ (dashed black lines), and temperature $T$ (solid orange lines) at times $t = 0$ (top) and $2 \times 10^6\,\mbox{yr}$ (bottom). As the overall surface density declines with time, the disk edge moves steadily outward. Most of the mass redistribution occurs within this first $10^6\,\mbox{yr}$, during which time the disk size increases to ${\sim}250\,\mbox{AU}$. Over a subsequent $10^6\,\mbox{yr}$ the disk size increases by only another ${\sim}50\,\mbox{AU}$. This late-time quiescent evolutionary phase is also characterized by values of $Q$ of order unity throughout the disk. Such behavior is consistent with that seen in more robust 2D simulations of disk evolution such as those performed by \citet{vorobyov2006,vorobyov2010}.

\subsection{Mass Accretion Rates}
\label{subsec:massaccretionrates}

\begin{figure}
\centering
\includegraphics[width=\columnwidth]{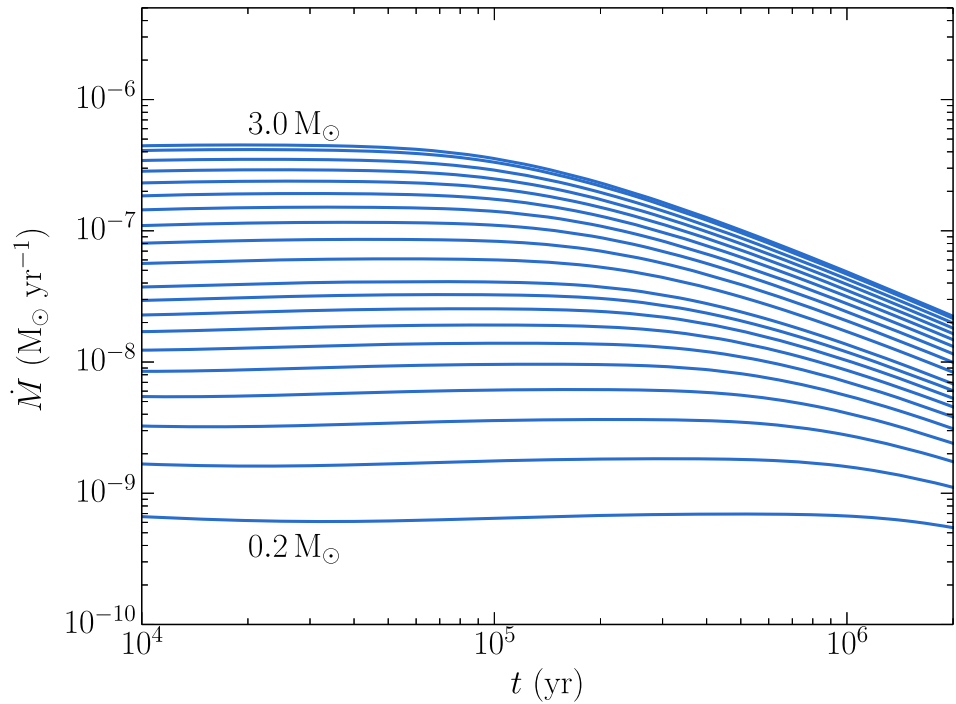}
\caption{Temporal evolution of the mass accretion rates for 21 protostars with initial masses with $0.2\,\mbox{M}_\odot \leq M_* \leq 3.0\,\mbox{M}_\odot$. The accretion rate for each protostar-disk system is initially relatively constant. Once the system settles into a quasi--steady state, the accretion rate declines as $t^{-6/5}$ \citep{lin1987}. The length of time any one system requires to reach the quasi--steady state scales roughly as $M_*^{-1/2}$. As a result, the difference in $\dot{M}$ between the least and most massive systems decreases with time.}
\label{fig:Mdot-t_toomreq}
\end{figure}

The instantaneous mass accretion rate onto the protostar is calculated from the change in the total disk mass between time steps, as mass loss from the disk occurs only through the inner boundary of the simulation domain (being that $r_{\rm edge}$ is always $\ll r_{\rm out}$). Figure \ref{fig:Mdot-t_toomreq} illustrates the time evolution of mass accretion rates between $10^4$ and $2{\times}10^6\,\mbox{yr}$ for protostars with initial masses of $0.2\,\mbox{M}_\odot \leq M_* \leq 3.0\,\mbox{M}_\odot$.

\begin{figure}
\centering
\includegraphics[width=\columnwidth]{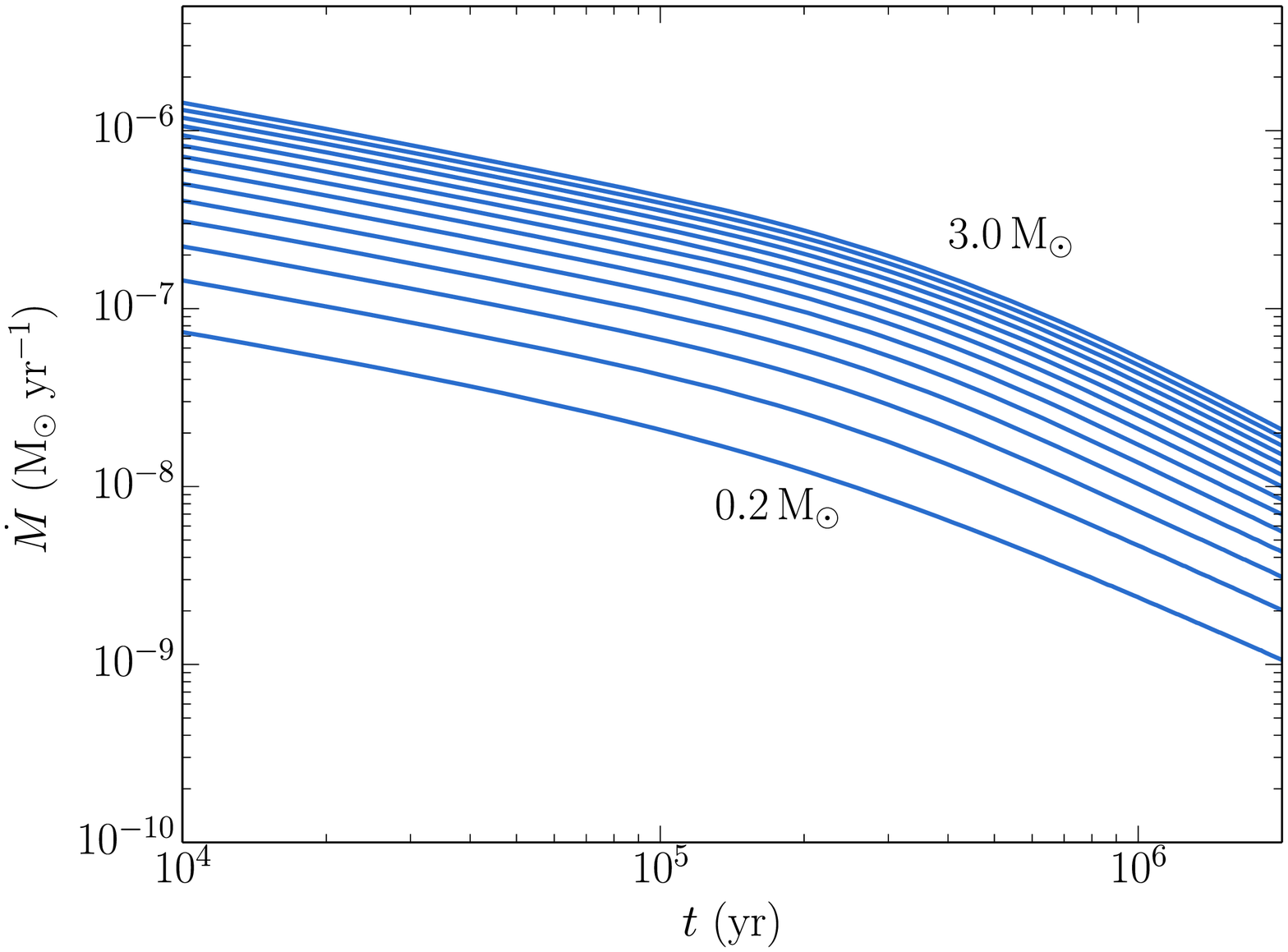}
\caption{Temporal evolution of the mass accretion rates for 21 protostars with initial masses with $0.2\,\mbox{M}_\odot \leq M_* \leq 3.0\,\mbox{M}_\odot$. For comparison, the disk evolution for these systems is governed by a standard $\alpha$ parameterization \citep[e.g.,][]{shakura1973}, with $\alpha = 10^{-2}$. There is an approximately one order of magnitude difference in $\dot{M}$ between the least and most massive protostar-disk systems. The constant value of this range of $\dot{M}$ on a logarithmic scale is in sharp contrast to the decrease in range seen in disks whose viscosity is described by gravitational torques.}
\label{fig:Mdot-t_alpha}
\end{figure}

The mass accretion rates onto the star are relatively constant at all masses during the first ${\sim}10^5\,\mbox{yr}$ as the disks approach a quasi--steady state, characterized by the radial mass accretion rate throughout the disk being uniform. The range in $\dot{M}$ between the least and most massive systems spans more than three orders of magnitude during this time: the $0.2\,\mbox{M}_\odot$ protostar accretes material from its disk at a rate of several times $10^{-10}\,\mbox{M}_\odot\,\mbox{yr}^{-1}$; the $3.0\,\mbox{M}_\odot$ protostar, at a rate of a few times $10^{-7}\,\mbox{M}_\odot\,\mbox{yr}^{-1}$.

Once a disk has settled into a quasi--steady state however, $\dot{M}$ begins to decline as $t^{-6/5}$ \citep[as found by][]{lin1987}. Transition into this regime is a direct consequence of the parameterization of the effective kinematic viscosity in terms of Toomre's $Q$ criterion---specifically the strong dependence of $\nu$ on $\Sigma$ in equation (\ref{eqn:convenientnu}). We find that the length of time preceding the transition scales approximately as $M_*^{-1/2}$. This causes the range of accretion rates spanned by systems with different initial masses to decrease with time.

For contrast, in Figure \ref{fig:Mdot-t_alpha} we provide an example of the analogous evolution of $\dot{M}$ for disks in which the effective kinematic viscosity is described by the spatio-temporally constant classical $\alpha$ parameterization of equation (\ref{eqn:alphaviscosity}) \citep{shakura1973}. We take $\alpha$ to have a fiduciary value of $10^{-2}$ in these models, consistent with estimates inferred from fitting disk similarity solutions to statistically significant samples of disk observations spanning different ages \citep[e.g.,][]{andrews2009}. The disks modeled in this fashion exhibit a steady decline in their mass accretion rates with time, at the same rate irrespective of the mass of the system. It is clear that for these objects there is no change in the range of $\dot{M}$ spanned by the least and most massive protostar-disk systems. The narrower (and constant in logarithmic space) range of accretion rates in these models makes it less likely that they can fit the observed $\dot{M}\mbox{--}M_*$ correlation.

\subsection{$\dot{M}\mbox{--}M_*$ Correlation}
\label{subsec:mdotmcorrelation}

\begin{figure}
\centering
\includegraphics[width=\columnwidth]{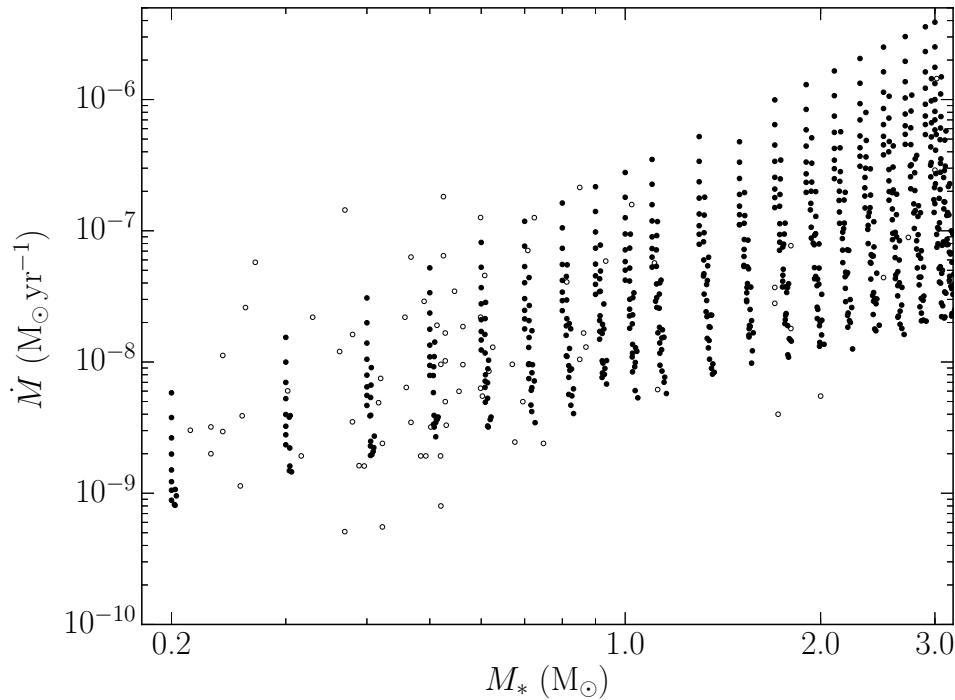}
\caption{Mass accretion rates $\dot{M}$ versus protostellar mass $M_*$. Open circles are observational measurements of objects with mass $\ge 0.1\,\mbox{M}_\odot$ from \citet[][and references therein]{muzerolle2005}. We evolve more than two hundred different initial protostar-disk configurations over $2{\times}10^6\,\mbox{yr}$. Filled circles are the mass accretion rates for these systems at every 1,000th time step. The orange line is the least squares fit to the observed mass accretion rates, and produces a $\beta = 1.3 \pm 0.3$. The blue line is the equivalent fit to the values of $\dot{M}$ from our simulations, and produces a $\beta = 1.7$.}
\label{fig:Mdot-M_composite}
\end{figure}

For the regime of intermediate to upper mass T Tauri stars ($M_* > 0.2\,\mbox{M}_\odot$), the exponent of the power law correlation $\beta$, between mass accretion rate $\dot{M}$ and protostellar mass $M_*$, can be taken to be approximately $1.3$ \citep[e.g.,][]{muzerolle2005,herczeg2008}. In Figure \ref{fig:Mdot-M_composite} we present mass accretion rates from more than 200 individual simulations, which reflect the evolution of protostars with masses of $0.2\,\mbox{M}_\odot \le M_* \le 3.0\,\mbox{M}_\odot$, and their disks, over $2{\times}10^6\,\mbox{yr}$. Although material is being accreted from the disk and onto the protostar in these simulations, the change in $M_*$ with time is negligible compared to the order of magnitude changes in $\dot{M}$ over the same period. A single simulation thus produces a seemingly vertical evolutionary track within the figure. For clarity, we plot only those values of $\dot{M}$ at every 1,000th time step from the individual simulations. Variations in the initial disk size---$r_{\rm edge}$, as determined through equation (\ref{eqn:diskradiusrelation})---cause protostar-disk systems with the same initial mass to follow slightly different evolutionary trajectories. The open circles are observational measurements of $\dot{M}$ for protostars in the same mass range as those of our simulation, from the compilation of \citet{muzerolle2005}. A least squares fit to this data produces $\beta = 1.3 \pm 0.3$. 

If we were to consider isochrones connecting together systems of different mass at the same age in Figure \ref{fig:Mdot-M_composite}, we would see that $\beta$ decreases as a function of protostellar age. {\bf This is due to the faster convergence to the self-similar solution for models of higher mass. The effect is} most apparent in the difference in the slope of the lines that form the upper and lower envelope bounding the simulation data at $t = 0$ and $2{\times}10^6\,\mbox{yr}$, respectively. Note that only a handful of the observational measurements fall outside of this envelope. The most significant outliers are the particularly high accreters found at the lowest masses. However, we suspect these objects are likely younger than those represented in our simulations, and still enshrouded by the material of their natal environment. Infall from their surroundings can induce disk instability and fragmentation. The resulting burst mode of accretion \cite[e.g.,][]{vorobyov2006,vorobyov2010} is known to dominate the quiescent mode of accretion (modeled here) at early times. Another challenge to our models comes from observations of the Orion Nebula Cluster by \citet{manara2012}, who find that the highest mass objects have a longer time scale for decrease of $\dot{M}$ than do lower mass objects, so that the $\dot{M}\mbox{--}M_*$ relation becomes steeper at later times. This is the opposite trend of our models. It could be that the accretion of mass on to the disk (not a part of our model) is significant in the more massive systems, and that their evolution to the self-similar solution is then delayed in comparison to low mass disks. This effect is outside the scope of our current model, but is a subject for future work.

\begin{figure}
\centering
\includegraphics[width=\columnwidth]{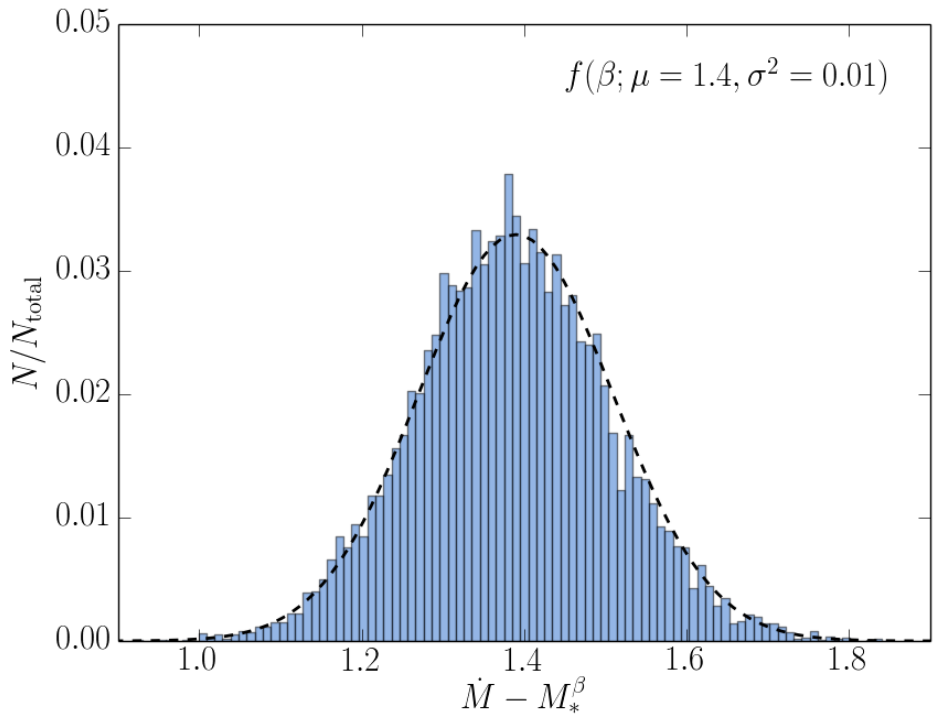}
\caption{A histogram of the slopes $\beta$ for the correlation between mass accretion rate and protostellar mass, $\dot{M}\mbox{--}M_*^\beta$, for 10,000 samples generated by method described in the text. The distribution of slopes is reasonably well fit by a Gaussian curve with mean $\mu = 1.4$ and variance $\sigma^2 = 0.01$ (dashed black line).}
\label{fig:slopehistogram}
\end{figure}

In order to make a fit of the $\dot{M}\mbox{--}M_*$ from our model data, we need to consider the relative weighting of models of different mass. To explore this possibility we randomly sample the spectrum of $\dot{M}\mbox{--}M_*$ results from our simulations using an initial mass function (IMF) of the form proposed by \citet{chabrier2005proc}. Initial protostellar masses are acquired from the Chabrier IMF. From these we determine a disk mass $M_{\rm disk} \propto M_*^{0.3}$ (as in Section \ref{subsec:initialconditions}) and size $r_{\rm edge}$ (equation \ref{eqn:diskradiusrelation}) at time $t = 0$ for each model. We then uniformly sample each model's temporal history in order to determine a specific value of $\dot{M}$. After 100 such ``measurements'' we are then able to estimate a value of $\beta$. Figure \ref{fig:slopehistogram} presents a histogram summarizing the value of $\beta$ from 10,000 such samplings. The average value of the exponent of the power law correlation in $\dot{M}\mbox{--}M_*$ is $\beta = 1.4$, with a standard deviation of ${\sim}0.1$. This places the value of $\beta$ as determined from our simulations well within the error bounds of the observationally determined value {\bf $1.3\, \pm\,0.3$}.

\begin{figure}
\centering
\includegraphics[width=\columnwidth]{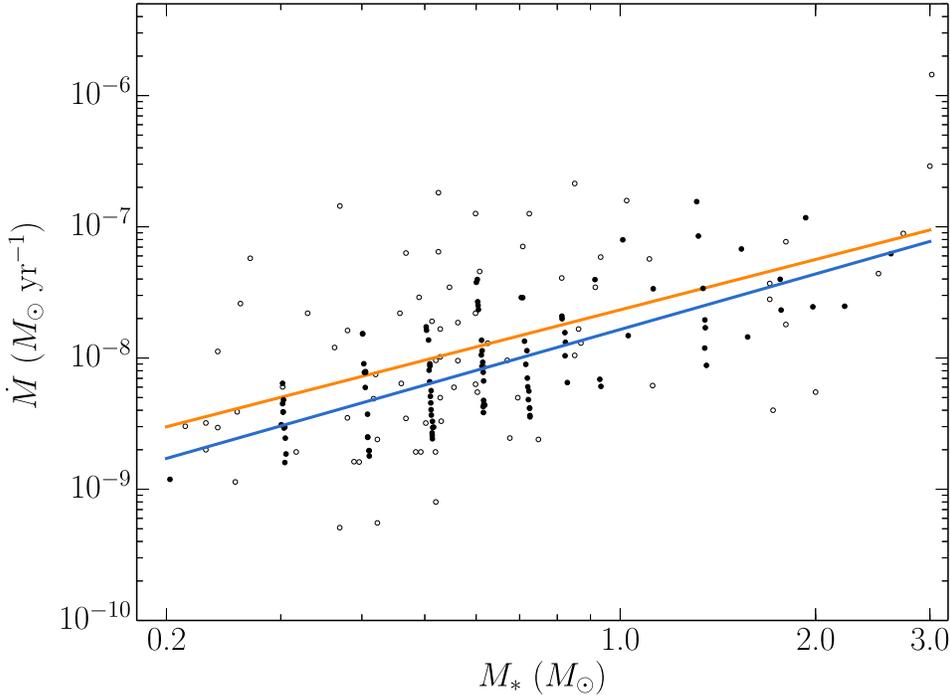}
\caption{We highlight one example of a plot of $\dot{M}\mbox{--}M_*$ generated by the selection criterion discussed in the text. For reference, the open circles are the observational measurements of T Tauri stars in the mass range of $0.2\,\mbox{M}_\odot < M_* < 3.0\,\mbox{M}_\odot$, from \citet[][and references therein]{muzerolle2005}. The orange line is the least squares fit to the observational data, and yields a value for the power law exponent of $\beta = 1.3\,\pm\,0.3$. The filled circles are the data obtained from our simulations, and the blue line is the least squares fit to these points, yielding $\beta = 1.4\,\pm\,0.1$.}
\label{fig:syntheticMdot-M}
\end{figure}

\begin{figure}
\centering
\includegraphics[width=\columnwidth]{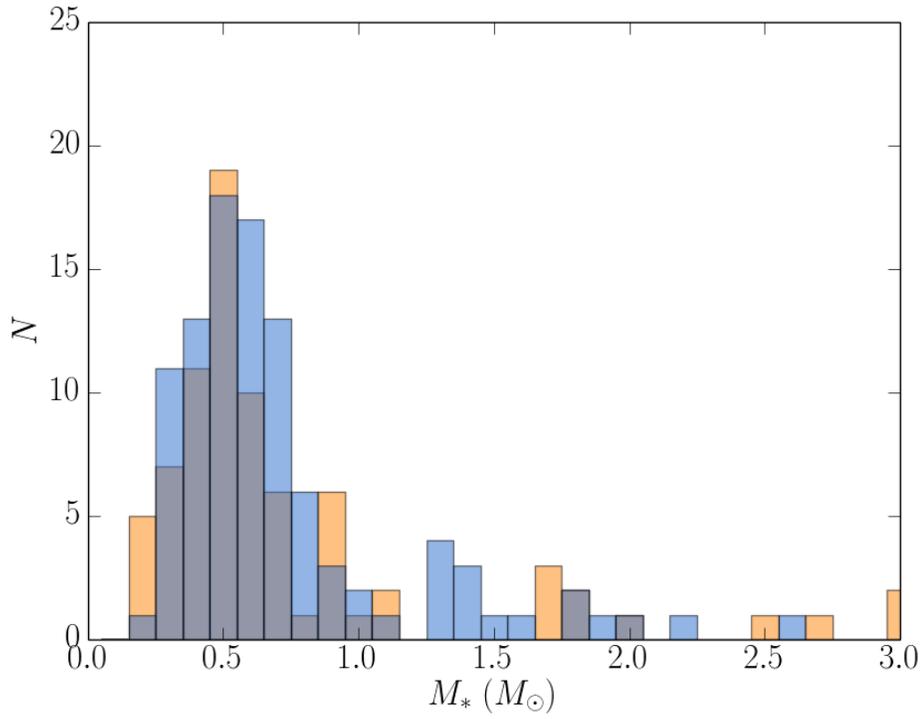}
\includegraphics[width=\columnwidth]{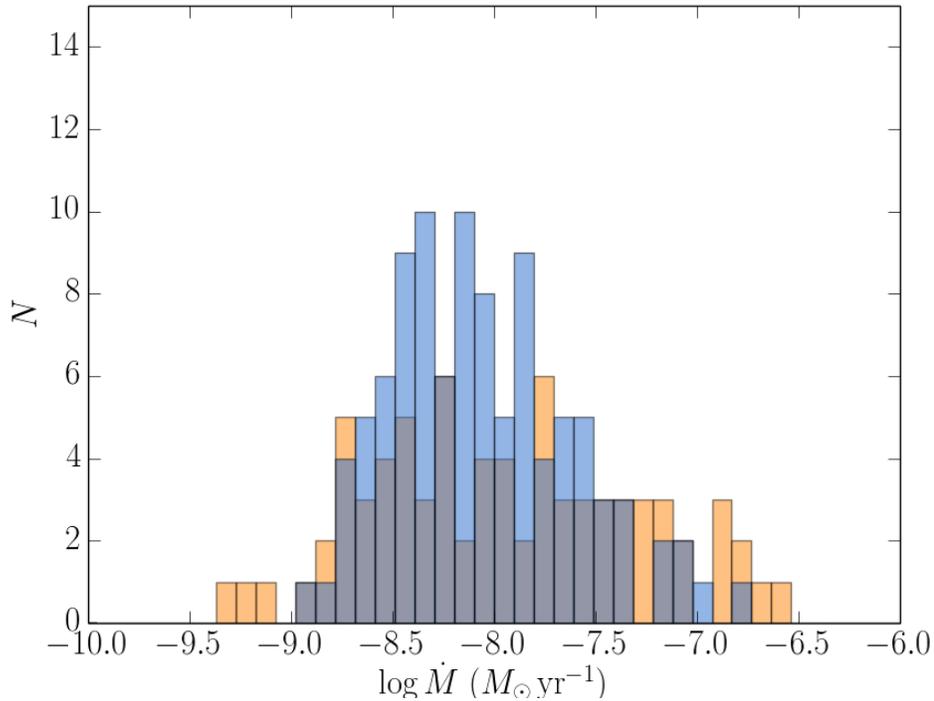}
\caption{Histograms of the number distributions by protostellar mass $M_*$ (top), and by mass accretion rate $\dot{M}$ (bottom), of both the observational measurements (orange) and simulation data (blue), gray regions indicating overlap, as they appear in Figure \ref{fig:syntheticMdot-M}. A Kolmogorov-Smirnov two-sample test, comparing the number distributions of the observational measurements to their corresponding simulation counterparts, produces a $p$ value of roughly $0.10$ in each case. The randomly selected simulation data is thus statistically indistinguishable from the observational measurements.}
\label{fig:numberdistributions}
\end{figure}

In Figure \ref{fig:syntheticMdot-M} we present one example of this sampling procedure for which the simulation data produces a typical value for the power law exponent of $\beta = 1.4\,\pm\,0.1$. In Figure \ref{fig:numberdistributions} we also present the number distributions of protostellar masses, and of mass accretion rates, for the observational measurements and the randomly selected simulation points. By visual inspection, the observational measurements and the sample drawn from the simulations are in general agreement. However, we can quantify this agreement statistically, as well as across repeated samples, to evaluate the likelihood with which our model is capable of reproducing the observed correlation.

\begin{figure}
\centering
\includegraphics[width=\columnwidth]{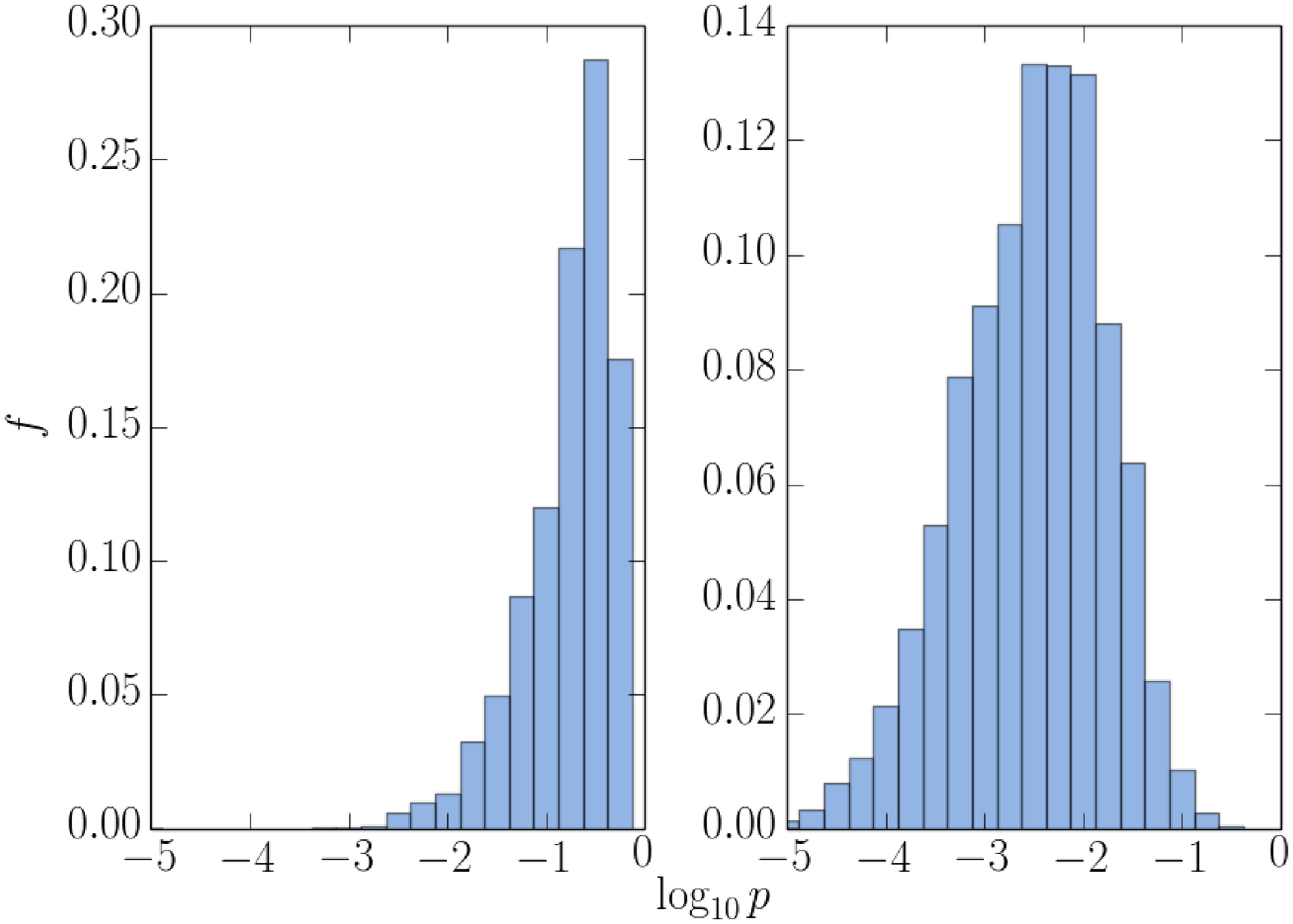}
\caption{
We perform 10,000 independent random samplings of the simulation data in the same manner described and depicted in Figure \ref{fig:numberdistributions}. For each sample we determine the number distribution by protostellar mass $M_*$ and mass accretion rate $\dot{M}$ and compare those distributions to their observational counterparts using a Kolmogorov-Smirnov two-sample test. The $p$ value resulting from each of these tests is depicted above, with $f$ denoting the fraction of the 10,000 simulations in the indicated bucket. Comparisons between these number distributions in $M_*$ appear at left, and in $\dot{M}$ at right.
}
\label{fig:pvaluedistributions}
\end{figure}

For each of the 10,000 samplings represented in Figure \ref{fig:slopehistogram}, we construct the corresponding number distributions by mass and mass accretion rate, as in Figure \ref{fig:numberdistributions}. We then perform a two-sample Kolmogorov-Smirnov test to evaluate the null hypothesis that the number distributions by mass and mass accretion rate between the observed and our randomly generated samples have been drawn from the same underlying distributions. Figure \ref{fig:pvaluedistributions} presents two histograms that summarize the $p$ values resulting from this test, where the $p$ value is an estimate of the probability that the two distributions are representative of a singular underlying sample; considered unlikely for $p < 0.05 \sim 2\sigma$. In general, we find $p$ to be sufficiently large such that the null hypothesis cannot be strictly ruled out. However, in roughly half of our randomly generated samples, we are unable to reproduce the magnitude spread in mass accretion rates that exists among observational measurements. As we discussed earlier, this is likely the result of younger objects than those represented in our simulations, in which infall from the surrounding environment is likely to dominate the accretion mode.

\section{Summary \& Discussion}
\label{sec:dnc}

In this paper, we have shown that the observed power law correlation between mass accretion rate $\dot{M}$ and protostellar mass $M_*$ can be explained within the framework of gravitational--torque--driven transport. We parameterize the effects of the gravitational torques as an effective kinematic viscosity using Toomre's $Q$ criterion \citep{toomre1964}, noting that this prescription resembles but also differs from the classical $\alpha$ model of \citet{shakura1973}.

We carry out more than 200 individual simulations of protostellar disks in order to examine the time evolution of their mass accretion rates in the $\dot{M}-M_*$ plane. The rates associated with a particular protostellar mass agree with those inferred from observational studies of T Tauri disks across a broad spectrum of protostellar masses. The observed scatter in $\dot{M}$ arises naturally as a result of the temporal evolution of the protostar-disk system through this plane. We are able to use a simple statistical argument, resampling our simulations onto the initial mass function of \citet{chabrier2005proc}, to show that even with limited sampling, our simulation results are sufficiently robust to be able to reproduce the observed correlation.

The initial disk masses presented in this paper are somewhat greater than is often reported in the literature \citep[by a factor of ${\sim}10$, e.g.,][]{andrews2005}. However, current estimates for disk masses based on dust emission may have been systematically underestimated \citep{hartmann2006,dunham2014}. Nevertheless, as the efficacy of our transport mechanism is dependent on disk mass, it is possible that that additional physics may be required at late times to remove the remaining disk material within observed disk lifetimes \citep{hernandez2008}. Conversely, there is growing evidence that some disks may persist for several Myr and have noticeable spiral structure, arcs, and gaps that may be indicative of gravitational instability \citep{liu2016}. One way to understand this is by invoking gas accretion on to the disks for an extended period of several Myr \citep{vorobyov2015b,lesur2015}. Accretion on to the disk is not a part of our current model and is a subject of future work.

In a recent paper, \citet{ercolano2014} offer a physically different model explanation. In their view, $\dot{M}$ is initially mass independent, and declines in a self-similar manner with a somewhat different power law index than in our model, due to the use of an $\alpha$-viscosity. Mass accretion is then quenched when a model wind mass loss rate (that depends on the X-ray luminosity of the protostar) equals the mass accretion rate. The physical view here is that the observed mass-dependent X-ray luminosity sets the $\dot{M}\mbox{--}M_*$ relation by reducing the accretion rate when it drops to the level of the wind mass loss rate. Mathematically however, both models depend partially upon a bimodal mass accretion rate history. It is only the physical explanation of the two phases, and the specific mathematical shape of their curves, that differ. The observed correlation in $\dot{M}\mbox{--}M_*$ may therefore be fit by a variety of models that have common mathematical elements, but differ substantially enough in their physics that they will hopefully lead to interesting observational comparator tests in the future.


\section*{Acknowledgements}
\label{sec:acknowledgements}

We thank the anonymous referee for comments that have significantly improved the discussion in the paper. SB acknowledges support from a Discovery Grant from the Natural Sciences and Engineering Research Council (NSERC) of Canada. This research has made use of NASA's Astrophysics Data System.


\begin{thebibliography}{65}
\expandafter\ifx\csname natexlab\endcsname\relax\def\natexlab#1{#1}\fi

\bibitem[{{Alexander} \& {Armitage}(2006)}]{alexander2006}
{Alexander}, R.~D. \& {Armitage}, P.~J. 2006, ApJL, 639, L83

\bibitem[{{Andr{\'{e}}} {et~al.}(1993){Andr{\'{e}}}, {Ward-Thompson}, \&
  {Barsony}}]{andre1993}
{Andr{\'{e}}}, P., {Ward-Thompson}, D., \& {Barsony}, M. 1993, ApJ, 406, 122

\bibitem[{{Andrews} \& {Williams}(2005)}]{andrews2005}
{Andrews}, S.~M. \& {Williams}, J.~P. 2005, ApJ, 631, 1134

\bibitem[{{Andrews} {et~al.}(2009){Andrews}, {Wilner}, {Hughes}, {Qi}, \&
  {Dullemond}}]{andrews2009}
{Andrews}, S.~M., {Wilner}, D.~J., {Hughes}, A.~M., {Qi}, C., \& {Dullemond},
  C.~P. 2009, ApJ, 700, 1502

\bibitem[{{Armitage} {et~al.}(2001){Armitage}, {Livio}, \&
  {Pringle}}]{armitage2001}
{Armitage}, P.~J., {Livio}, M., \& {Pringle}, J.~E. 2001, MNRAS, 324, 705

\bibitem[{{Balbus} \& {Hawley}(1991)}]{balbus1991}
{Balbus}, S.~A. \& {Hawley}, J.~F. 1991, ApJ, 376, 214

\bibitem[{{Blaes} \& {Hawley}(1991)}]{blaes1994}
Blaes, O.~M. \& Balbus, S.~A. 1994, ApJ, 421, 163

\bibitem[{{Basu}(1997)}]{basu1997}
{Basu}, S. 1997, ApJ, 485, 240

\bibitem[{{Beckwith} {et~al.}(1990){Beckwith}, {Sargent}, {Chini}, \&
  {Guesten}}]{beckwith1990}
{Beckwith}, S.~V.~W., {Sargent}, A.~I., {Chini}, R.~S., \& {Guesten}, R. 1990,
  ApJ, 99, 924

\bibitem[{{Binney} \& {Tremaine}(2008)}]{binney2008book}
{Binney}, J. \& {Tremaine}, S. 2008, {Galactic Dynamics} (Princeton University
  Press)

\bibitem[{{Boley} {et~al.}(2006){Boley}, {Mej{\'{\i}}a}, {Durisen}, {Cai},
  {Pickett}, \& {D'Alessio}}]{boley2006}
{Boley}, A.~C., {Mej{\'{\i}}a}, A.~C., {Durisen}, R.~H., {Cai}, K., {Pickett},
  M.~K., \& {D'Alessio}, P. 2006, ApJ, 651, 517

\bibitem[{{Brandenburg} {et~al.}(1995){Brandenburg}, {Nordlund}, {Stein}, \&
  {Torkelsson}}]{brandenburg1995}
{Brandenburg}, A., {Nordlund}, A., {Stein}, R.~F., \& {Torkelsson}, U. 1995,
  ApJ, 446, 741

\bibitem[{{Chabrier}(2005)}]{chabrier2005proc}
{Chabrier}, G. 2005, in The Initial Mass Function 50 Years Later, ed.
  E.~{Corbelli}, F.~{Palla}, \& H.~{Zinnecker}, 41

\bibitem[{{Cleeves} {et~al.}(2013){Cleeves}, {Adams}, \&
  {Bergin}}]{cleeves2013}
Cleeves, L.~I., Adams, F.~C., \& Bergin, E.~A. 2013, ApJ, 772, 5

\bibitem[{{Cossins} {et~al.}(2009){Cossins}, {Lodato}, \&
  {Clarke}}]{cossins2009}
{Cossins}, P., {Lodato}, G., \& {Clarke}, C.~J. 2009, MNRAS, 393, 1157

\bibitem[{{Dong} {et~al.}(2016){Dong}, {Vorobyov}, {Pavlyuchenkov}, {Chiang}, \& {Liu}}]{dong2016}
  {Dong}, R., {Vorobyov}, E., {Pavlyuchenkov}, Y., {Chiang}, E., \& {Liu}, H. B. 2016, 
  ApJ, 823, 141

\bibitem[{{D'Alessio} {et~al.}(2001){D'Alessio}, {Calvet}, \&
  {Hartmann}}]{d'alessio2001}
{D'Alessio}, P., {Calvet}, N., \& {Hartmann}, L. 2001, ApJ, 553, 321

\bibitem[{{Dullemond} {et~al.}(2006){Dullemond}, {Natta}, \&
  {Testi}}]{dullemond2006}
{Dullemond}, C.~P., {Natta}, A., \& {Testi}, L. 2006, ApJL, 645, L69

\bibitem[{{Dunham} {et~al.}(2014){Dunham}, {Vorobyov}, \&
  {Arce}}]{dunham2014}
{Dunham}, M.~M., {Vorobyov}, E.~I., \& {Arce}, H.~G. 2014, MNRAS, 444, 887

\bibitem[{{Eisner} \& {Carpenter}(2006)}]{eisner2006}
{Eisner}, J.~A. \& {Carpenter}, J.~M. 2006, ApJ, 641, 1162

\bibitem[{{Ercolano} {et~al.}(2014){Ercolano}, {Mayr}, {Owen}, {Rosotti}, \&
  {Manara}}]{ercolano2014}
{Ercolano}, B., {Mayr}, D., {Owen}, J.~E., {Rosotti}, G., \& {Manara}, C.~F.
  2014, MNRAS, 439, 256

\bibitem[{{Fleming} \& {Stone}(2003)}]{fleming2003}
{Fleming}, T. \& {Stone}, J.~M. 2003, ApJ, 585, 908

\bibitem[{{Fromang} {et~al.}(2002){Fromang}, {Terquem}, \&
  {Balbus}}]{fromang2002}
{Fromang}, S., {Terquem}, C., \& {Balbus}, S.~A. 2002, MNRAS, 329, 18

\bibitem[Fukagawa et al.(2004)]{fukagawa2004} Fukagawa, M., Hayashi, M., Tamura, M., et al.\ 2004, ApJ, 605, L53 

\bibitem[{{Gammie}(1996)}]{gammie1996}
{Gammie}, C.~F. 1996, ApJ, 457, 355

\bibitem[{{Gammie}(2001)}]{gammie2001}
{Gammie}, C.~F. 2001, ApJ, 553, 174

\bibitem[{{Garcia} {et~al.}(2001){Garcia}, {Cabrit}, {Ferreira}, \&
  {Binette}}]{garcia2001}
{Garcia}, P.~J.~V., {Cabrit}, S., {Ferreira}, J., \& {Binette}, L. 2001, A\&A,
  377, 609

\bibitem[Grady et al.(2001)]{grady2001} Grady, C.~A., Polomski, E.~F., Henning, T., et al.\ 2001, AJ, 122, 3396 

\bibitem[Grady et al.(2013)]{grady2013} Grady, C.~A., Muto, T., Hashimoto, J., et al.\ 2013, ApJ, 762, 48 

\bibitem[{{Hartmann} {et~al.}(1998){Hartmann}, {Calvet}, {Gullbring}, \&
  {D'Alessio}}]{hartmann1998}
{Hartmann}, L., {Calvet}, N., {Gullbring}, E., \& {D'Alessio}, P. 1998, ApJ,
  495, 385

\bibitem[{{Hartmann} {et~al.}(2006){Hartmann}, {D'Alessio}, {Calvet}, \&
  {Muzerolle}}]{hartmann2006}
{Hartmann}, L., {D'Alessio}, P., {Calvet}, N., \& {Muzerolle}, J. 2006, ApJ,
  648, 484

\bibitem[Hashimoto et al.(2011)]{hashimoto2011} Hashimoto, J., Tamura, M., Muto, T., et al.\ 2011, ApJ, 729, L17 

\bibitem[{{Hawley} {et~al.}(1996){Hawley}, {Gammie}, \& {Balbus}}]{hawley1996}
{Hawley}, J.~F., {Gammie}, C.~F., \& {Balbus}, S.~A. 1996, ApJ, 464, 690

\bibitem[{{Hayashi}(1981)}]{hayashi1981}
{Hayashi}, C. 1981, Progress of Theoretical Physics Supplement, 70, 35

\bibitem[{{Herczeg} \& {Hillenbrand}(2008)}]{herczeg2008}
{Herczeg}, G.~J. \& {Hillenbrand}, L.~A. 2008, ApJ, 681, 594

\bibitem[{{Hern{\'a}ndez} {et~al.}(2008){Hern{\'a}ndez}, {Hartmann}, {Calvet},
  {Jeffries}, {Gutermuth}, {Muzerolle}, \& {Stauffer}}]{hernandez2008}
{Hern{\'a}ndez}, J., {Hartmann}, L., {Calvet}, N., {Jeffries}, R.~D.,
  {Gutermuth}, R., {Muzerolle}, J., \& {Stauffer}, J. 2008, ApJ, 686, 1195

\bibitem[{{Hueso} \& {Guillot}(2005)}]{hueso2005}
{Hueso}, R. \& {Guillot}, T. 2005, A\&A, 442, 703

\bibitem[{{Igea} \& {Glassgold}(1999)}]{igea1999}
Igea, J. \& Glassgold, A. E. 1999, ApJ, 518, 848

\bibitem[{{Krasnopolsky} {et~al.}(2003){Krasnopolsky}, {Li}, \&
  {Blandford}}]{krasnopolsky2003}
{Krasnopolsky}, R., {Li}, Z.-Y., \& {Blandford}, R.~D. 2003, ApJ, 595, 631

\bibitem[{{Lada} \& {Wilking}(1984)}]{lada1984}
{Lada}, C.~J. \& {Wilking}, B.~A. 1984, ApJ, 287, 610

\bibitem[{{Larson}(2003)}]{larson2003}
{Larson}, R.~B. 2003, RPPh, 66, 1651

\bibitem[Lesur et al.(2015)]{lesur2015} Lesur, G., Hennebelle, P., \& Fromang, S.\ 2015, A\&A, 582, L9 

\bibitem[{{Lin} \& {Pringle}(1987)}]{lin1987}
{Lin}, D.~N.~C. \& {Pringle}, J.~E. 1987, MNRAS, 225, 607

\bibitem[Liu et al.(2016)]{liu2016} Liu, H.~B., Takami, M., Kudo, T., et al.\ 2016, Science Advances, 2, e1500875 

\bibitem[{{Lodato} \& {Rice}(2004)}]{lodato2004}
{Lodato}, G. \& {Rice}, W.~K.~M. 2004, MNRAS, 351, 630

\bibitem[{{Lynden-Bell} \& {Pringle}(1974)}]{lyndenbell1974}
{Lynden-Bell}, D. \& {Pringle}, J.~E. 1974, MNRAS, 168, 603

\bibitem[Manara et al.(2012)]{manara2012} Manara, C.~F., Robberto, M., Da Rio, N., et al.\ 2012, ApJ, 755, 154 

\bibitem[{{McCaughrean} \& {O'Dell}(1996)}]{mccaughrean1996}
{McCaughrean}, M.~J. \& {O'Dell}, C.~R. 1996, AJ, 111, 1977

\bibitem[Muto et al.(2012)]{muto2012} Muto, T., Grady, C.~A., Hashimoto, J., et al.\ 2012, ApJ, 748, L22 

\bibitem[{{Muzerolle} {et~al.}(2005){Muzerolle}, {Luhman}, {Brice{\~n}o},
  {Hartmann}, \& {Calvet}}]{muzerolle2005}
{Muzerolle}, J., {Luhman}, K.~L., {Brice{\~n}o}, C., {Hartmann}, L., \&
  {Calvet}, N. 2005, ApJ, 625, 906

\bibitem[{{O'Dell} \& {Wen}(1994)}]{o'dell1994}
{O'Dell}, C.~R. \& {Wen}, Z. 1994, ApJ, 436, 194

\bibitem[{{Pringle}(1981)}]{pringle1981}
{Pringle}, J.~E. 1981, ARA\&A, 19, 137

\bibitem[{{Rice} \& {Armitage}(2009)}]{rice2009}
{Rice}, W.~K.~M. \& {Armitage}, P.~J. 2009, MNRAS, 396, 2228

\bibitem[{{Rigliaco} {et~al.}(2011){Rigliaco}, {Natta}, {Randich}, {Testi}, \&
  {Biazzo}}]{rigliaco2011}
{Rigliaco}, E., {Natta}, A., {Randich}, S., {Testi}, L., \& {Biazzo}, K. 2011,
  A\&A, 525, A47

\bibitem[{{Sano} {et~al.}(2000)}]{sano2000}
Sano, T., Miyama, S.~M., Umebayashi, T., \& Nakano, T. 2000, ApJ, 543, 486

\bibitem[{{Shakura} \& {Sunyaev}(1973)}]{shakura1973}
{Shakura}, N.~I. \& {Sunyaev}, R.~A. 1973, A\&A, 24, 337

\bibitem[{{Shampine}(1994)}]{shampine1994book}
{Shampine}, L.~E. 1994, Numerical Solution of Ordinary Differential Equations
  (Chapman \& Hall)

\bibitem[{{Shu} {et~al.}(1987){Shu}, {Adams}, \& {Lizano}}]{shu1987}
{Shu}, F.~H., {Adams}, F.~C., \& {Lizano}, S. 1987, ARA\&A, 25, 23

\bibitem[{{Stone} {et~al.}(2000){Stone}, {Gammie}, {Balbus}, \&
  {Hawley}}]{pp4stone2000coll}
{Stone}, J.~M., {Gammie}, C.~F., {Balbus}, S.~A., \& {Hawley}, J.~F. 2000, in
  Protostars and Planets IV, ed. {{Mannings}, V. and {Boss}, A.~P. and
  {Russell}, S.~S.} (University of Arizona Press)

\bibitem[Tamura(2016)]{tamura2016} Tamura, M.\ 2016, Proceeding of the Japan Academy, Series B, 92, 45 

\bibitem[{{Terebey} {et~al.}(1984){Terebey}, {Shu}, \& {Cassen}}]{terebey1984}
{Terebey}, S., {Shu}, F.~H., \& {Cassen}, P. 1984, ApJ, 286, 529

\bibitem[{{Toomre}(1964)}]{toomre1964}
{Toomre}, A. 1964, ApJ, 139, 1217

\bibitem[{{Vorobyov}(2011)}]{vorobyov2011}
{Vorobyov}, E.~I. 2011, ApJ, 729, 146

\bibitem[{{Vorobyov} \& {Basu}(2005{\natexlab{a}})}]{vorobyov2005a}
{Vorobyov}, E.~I. \& {Basu}, S. 2005{\natexlab{a}}, MNRAS, 360, 675

\bibitem[{{Vorobyov} \& {Basu}(2005{\natexlab{b}})}]{vorobyov2005b}
---. 2005{\natexlab{b}}, ApJL, 633, L137

\bibitem[{{Vorobyov} \& {Basu}(2006)}]{vorobyov2006}
---. 2006, ApJ, 650, 956

\bibitem[{{Vorobyov} \& {Basu}(2007)}]{vorobyov2007}
---. 2007, MNRAS, 381, 1009

\bibitem[{{Vorobyov} \& {Basu}(2008)}]{vorobyov2008}
---. 2008, ApJL, 676, L139

\bibitem[{{Vorobyov} \& {Basu}(2009{\natexlab{a}})}]{vorobyov2009a}
---. 2009{\natexlab{a}}, MNRAS, 393, 822

\bibitem[{{Vorobyov} \& {Basu}(2009{\natexlab{b}})}]{vorobyov2009b}
---. 2009{\natexlab{b}}, ApJ, 703, 922

\bibitem[{{Vorobyov} \& {Basu}(2010)}]{vorobyov2010}
---. 2010, ApJ, 719, 1896

\bibitem[{{Vorobyov} \& {Basu}(2015)}]{vorobyov2015}
---. 2015, ApJ, 805, 115

\bibitem[Vorobyov et al.(2015)]{vorobyov2015b} Vorobyov, E.~I., Lin, D.~N.~C., \& Guedel, M.\ 2015, A\&A, 573, A5 

\bibitem[{{Williams}(2012)}]{williams2012}
{Williams}, J.~P. 2012, M\&PS, 47, 1915

\bibitem[{{Williams} \& {Cieza}(2011)}]{williams2011}
{Williams}, J.~P. \& {Cieza}, L.~A. 2011, ARA\&A, 49, 67

\bibitem[{{Yorke} {et~al.}(1993){Yorke}, {Bodenheimer}, \&
  {Laughlin}}]{yorke1993}
{Yorke}, H.~W., {Bodenheimer}, P., \& {Laughlin}, G. 1993, ApJ, 411, 274

\bibitem[{{Zhu} {et~al.}(2009){Zhu}, {Hartmann}, {Gammie}, \&
  {McKinney}}]{zhu2009}
{Zhu}, Z., {Hartmann}, L., {Gammie}, C., \& {McKinney}, J.~C. 2009, ApJ, 701,
  620

\bibitem[{{Zhu} {et~al.}(2010){Zhu}, {Hartmann}, {Gammie}, {Book}, {Simon}, \&
  {Engelhard}}]{zhu2010}
{Zhu}, Z., {Hartmann}, L., {Gammie}, C.~F., {Book}, L.~G., {Simon}, J.~B., \&
  {Engelhard}, E. 2010, ApJ, 713, 1134

\end{thebibliography}


\end{document}